\documentclass[
    ,final            
  ]
  {aipproc}

\layoutstyle{6x9}

\newcommand{\lsim}{\,{\buildrel < \over {_\sim}}\,}

\newcommand{\sqrtsNN}{\sqrt{s_{\scriptscriptstyle{{\rm NN}}}}}

\newcommand{\gev}{\mathrm{GeV}}
\newcommand{\tev}{\mathrm{TeV}}
\newcommand{\fm}{\mathrm{fm}}

\newcommand{\RAA}{R_{\rm AA}}

\newcommand{\pt}{p_T}

\begin{document}

\title{Heavy-quark energy loss at RHIC and LHC}


\author{A.~Dainese}{
  address={Universit\`a degli Studi di Padova and INFN, Padova, Italy}
}

\author{N.~Armesto}{
  address={Dep. de F\'{\i}sica de Part\'{\i}culas and IGFAE,
Universidade de Santiago de Compostela, 
Spain}
}

\author{M.~Cacciari}{
  address={LPTHE, Universit\'e Pierre et Marie Curie (Paris 6), France}
}

\author{C.A.~Salgado}{
  address={Department of Physics, CERN, Theory Division,
   Gen\`eve, Switzerland}
}

\author{U.A.~Wiedemann$^\ddagger$}{
  address={Department of Physics and Astronomy, University of Stony Brook, 
  NY, USA}
}

\begin{abstract}
The attenuation of heavy-flavored particles in 
nucleus--nucleus collisions tests
the microscopic dynamics of medium-induced parton energy 
loss and, in particular, its expected dependence on the identity 
(color charge and mass) of the parent parton.
We discuss the comparison of theoretical calculations with 
recent single-electron data from RHIC experiments. Then, we present 
predictions for the 
heavy-to-light ratios of $D$ and $B$ mesons at LHC energy.
\end{abstract}

\maketitle


\section{Introduction}

Believed to be the main origin of the jet quenching phenomena 
observed~\cite{nagle} in 
nucleus--nucleus collisions at RHIC energy 
$\sqrtsNN=62$--$200~\gev$, 
parton energy loss via gluon-radiation is expected to depend
on the properties (gluon density and volume) of the `medium' formed in the 
collision  
and on the properties (color charge and mass) of the `probe' 
parton~\cite{wiedemann}.
Hard gluons would lose more energy than hard quarks due to the stronger 
color coupling with the medium.   
In addition, charm and beauty quarks are 
qualitatively different probes with respect to
 light partons, since their energy loss
is expected to be reduced, as a consequence of a mass-dependent restriction 
in the phase-space
into which gluon radiation can 
occur~\cite{dokshitzerkharzeev,aswmassive,djordjevic,zhang}.

We study quenching effects for heavy quarks by supplementing
perturbative QCD calculations
of the baseline $\pt$ distributions 
with in-medium energy loss, included via 
the BDMPS quenching weights. 
The quenching weights,
 computed for
light quarks and gluons in~\cite{sw} and for heavy quarks in~\cite{adsw}, 
depend on the 
transport coefficient $\hat{q}$, a measure of the medium density, 
and on the in-medium path length. These inputs are 
evaluated on a parton-by-parton level, using a Glauber-model based 
description of the local $\hat{q}$ profile in the transverse 
direction~\cite{pqm}. The $\hat{q}$ value 
is chosen in order to reproduce the light-flavor particles
nuclear modification factor $\RAA(\pt)$ measured in central 
\mbox{Au--Au} collisions at $\sqrtsNN=200~\gev$ (Fig.~\ref{fig:rhic}, left):
the range favored by the data for the
parton-averaged transport coefficient is $\hat{q}=4$--$14~\gev^2/\fm$. 

\section{Single electrons at RHIC}

\begin{figure}[!t]
 \begin{tabular}{lc}
  \begin{minipage}{0.41\linewidth}
  \includegraphics[width=\textwidth]{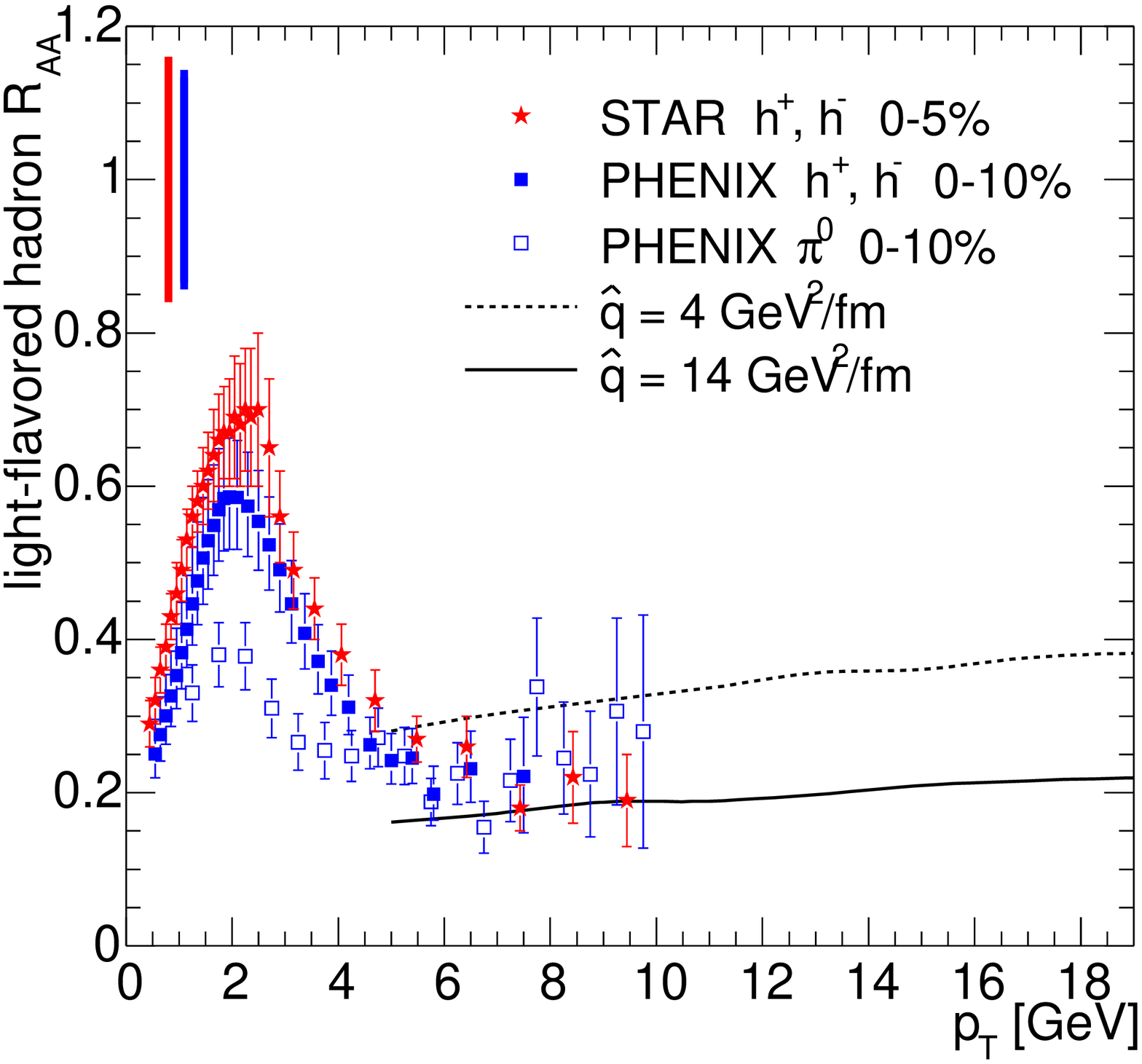}
  \end{minipage}
      &
  \begin{minipage}{0.59\linewidth}
  \includegraphics[angle=-90,width=.95\textwidth]{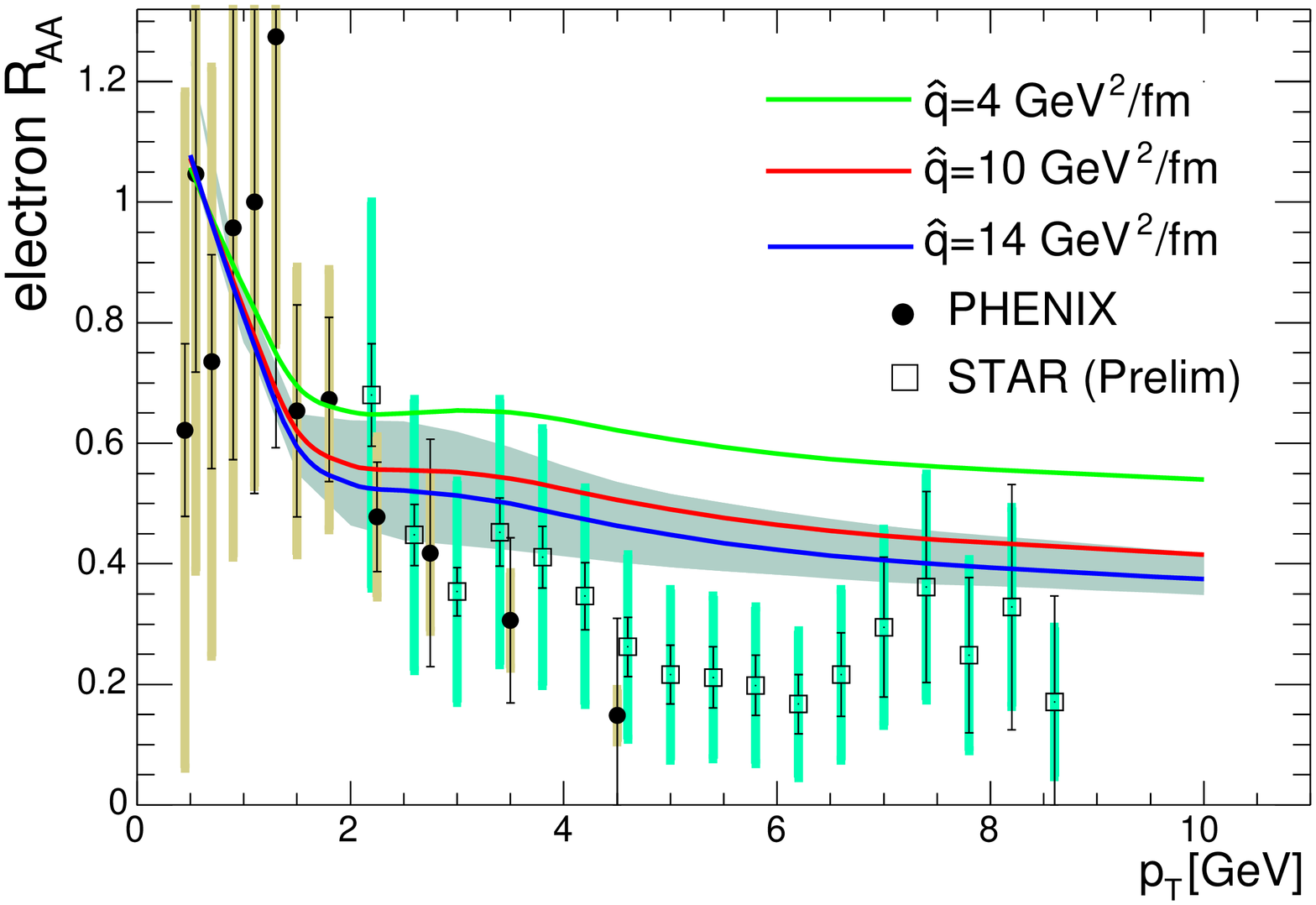}
  \end{minipage}
 \end{tabular}
 \caption{Central Au--Au collisions at $\sqrtsNN=200~\gev$: 
          $\RAA$ for light-flavored hadrons 
          (left, adapted from~\cite{pqm}) and for heavy-flavor decay electrons
          (right, from~\cite{acdsw})}
 \label{fig:rhic}
\end{figure}

Heavy-quark energy loss is presently studied at RHIC using
measurements of the nuclear modification factor $\RAA$ 
of `non-photonic' ($\gamma$-conversion- and $\pi^0$-Dalitz-subtracted) 
single electrons. The most recent data by PHENIX~\cite{phenixe} 
and STAR~\cite{stare}, reaching out 
to 5 and 9~GeV, respectively, are shown in Fig.~\ref{fig:rhic} (right). 
Since this is an inclusive measurement, with charm decays dominating at low
$\pt$ and beauty decays dominating at high $\pt$, the comparison with 
mass-dependent 
energy loss predictions should rely on a solid and data-validated 
pp baseline. Such baseline is still lacking at the moment, as we explain 
in the following.
The state-of-the-art perturbative predictions (FONLL), that we use
as a baseline, indicate that, in pp collisions, 
charm decays dominate the electron 
$\pt$ spectrum up to about 5~GeV~\cite{acdsw}. 
However, there is a large perturbative uncertainty on 
position in $\pt$ of the $c$-decay/$b$-decay 
crossing point: depending on the choice of the factorization 
and renormalization scales this position 
can vary from 3 to 9~GeV~\cite{acdsw}. 
In addition, the calculation tends to underpredict 
the non-photonic electron spectrum measured in pp collisions~\cite{acdsw}. 

For our electron $\RAA$ results (Fig.~\ref{fig:rhic}, right), 
in addition to the uncertainty on the medium density (curves for 
$\hat{q}=4$, 10, $14~\gev^2/\fm$), we also account for
the perturbative uncertainty by varying the values of the
scales and of the $c$ and $b$ quark masses (shaded band associated to 
the $\hat{q}=14~\gev^2/\fm$ curve)~\cite{acdsw}.
We find that the nuclear modification factor of single electrons is 
about 0.2 larger than that of light-flavor hadrons. Thus, electrons 
are in principle sensitive to the mass hierarchy of parton energy loss.
The available data neither allow us to support claims of inconsistency 
between theory and experiment, nor do they support yet the expected 
mass hierarchy.
It is important to note that, in general, the perturbative uncertainty in 
calculating the partonic baseline spectrum is comparable to the 
model-intrinsic uncertainty in determining $\hat{q}$.
If future experimental studies at RHIC succeeded in disentangling the 
charm and beauty contributions to single electrons, the sensitivity in the
theory-data comparison would be largely improved.

\section{Heavy-to-light ratios at LHC}

\begin{figure}[!t]
\includegraphics[width=0.85\textwidth]{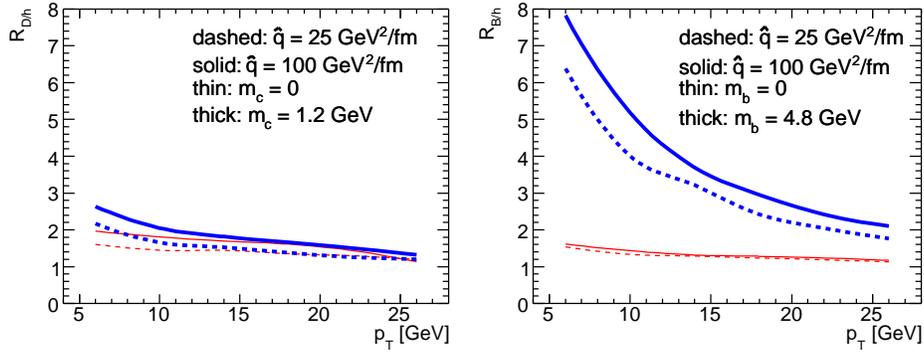}
\caption{Heavy-to-light ratios for $D$ (left) 
         and $B$ (right) mesons 
         for the case of realistic heavy-quark masses 
         and for a case
         study in which the quark mass dependence of parton energy 
         loss is neglected~\cite{adsw}}
\label{fig:lhc}
\end{figure}

Heavy quarks will be produced with large cross sections at LHC energy 
and the experiments will be equipped with detectors optimized for the
separation of charm and beauty decay vertices. Thus, it should be possible
to carry out a direct comparison of the attenuation of light-flavor
hadrons, $D$ mesons, and $B$ mesons. 
We calculate the expected nuclear modification factors $\RAA$ exploring 
a conservatively-large range in the medium density for central \mbox{Pb--Pb}
collisions at $\sqrtsNN=5.5~\tev$: $25<\hat{q}<100~\gev^2/\fm$. 
We use standard NLO
perturbative predictions for the $c$ and $b$ 
$\pt$-differential cross sections~\cite{mnr}.

Figure~\ref{fig:lhc} (thick lines) 
shows our results for the heavy-to-light ratios 
of $D$ and $B$ mesons~\cite{adsw}, 
defined as the ratios of the nuclear
modification factors of $D(B)$ mesons to that of light-flavor hadrons ($h$):
$R_{D(B)/h}=\RAA^{D(B)}/\RAA^h$. We illustrate the effect of the mass 
by artificially neglecting the  
mass dependence of parton energy loss (thin curves). 
The enhancement above unity that persists in the $m_{c(b)}=0$ 
cases is mainly due to the
color-charge dependence of energy loss,
since at LHC energy most of the light-flavor hadrons 
will originate from a gluon parent. Our results indicate that, for $D$ mesons,
the mass effect is small and 
limited the region $\pt\lsim 10~\gev$, while for $B$
mesons a large enhancement can be expected up to $20~\gev$.
Therefore, the comparison of the high-$\pt$ suppression
for $D$ mesons and for light-flavor hadrons will test the color-charge 
dependence (quark parent vs. gluon parent) of parton energy loss,
while the comparison for $B$ mesons and for light-flavor hadrons 
will test its mass dependence~\cite{adsw}.



\begin{thebibliography}{99}

\bibitem{nagle}
 J.~Nagle, {\em these proceedings}.

\bibitem{wiedemann}
 U.~A.~Wiedemann, {\em these proceedings}.

\bibitem{dokshitzerkharzeev}
  Yu.~L.~Dokshitzer and D.~E.~Kharzeev, {\em Phys.~Lett.}~{\bf B519}, 199 
  (2001).

\bibitem{aswmassive}
  N.~Armesto, C.~A.~Salgado and U.~A.~Wiedemann,
  {\em Phys.~Rev.}~{\bf D69}, 114003 (2004).

\bibitem{djordjevic}
  M.~Djordjevic and M.~Gyulassy,
  {\em Nucl.~Phys.}~{\bf A733}, 265 (2004).

\bibitem{zhang}
  B.~W.~Zhang, E.~Wang and X.~N.~Wang,
  {\em Phys.~Rev.~Lett.}~{\bf 93}, 072301 (2004).

\bibitem{sw}
  C.~A.~Salgado and U.~A.~Wiedemann,
  {\em Phys.~Rev.}~{\bf D68}, 014008 (2003).

\bibitem{adsw}
  N.~Armesto, A.~Dainese, C.~A.~Salgado and U.~A.~Wiedemann,
  {\em Phys.~Rev.}~{\bf D71}, 054027 (2005).

\bibitem{pqm}
  A.~Dainese, C.~Loizides and G.~Pai\'c,
  {\em Eur.~Phys.~J.}~{\bf C38}, 461 (2005).

\bibitem{phenixe}
  S.~S.~Adler {\it et al.}, PHENIX Collaboration, arXiv:nucl-ex/0510047. 

\bibitem{stare}
  J.~Bielcik {\it et al.}, STAR Collaboration, arXiv:nucl-ex/0511005. 

\bibitem{acdsw}
  N.~Armesto, M.~Cacciari, A.~Dainese, C.~A.~Salgado and U.~A.~Wiedemann,
  arXiv:hep-ph/0511257.

\bibitem{mnr} 
  M.~L.~Mangano, P.~Nason and G.~Ridolfi, 
  {\em Nucl.~Phys.}~{\bf B373}, 295 (1992).

\end{thebibliography}
\end{document}